\documentclass{emulateapj}

\shorttitle{Lyman-$\alpha$ Transit Spectroscopy of GJ436b}
\shortauthors{Kulow et al.}

\usepackage{graphics}
\usepackage{natbib}
\usepackage{epsfig}
\usepackage{amsmath}
\begin{document}

\title{Lyman-$\alpha$ Transit Spectroscopy and the Neutral Hydrogen Tail of the Hot Neptune GJ436b}

\author{Jennifer R. Kulow\altaffilmark{1}, Kevin France\altaffilmark{1,3}, Jeffery Linsky\altaffilmark{2}, and R. O. Parke Loyd\altaffilmark{1}}
\altaffiltext{1}{Center for Astrophysics and Space Astronomy, University of Colorado, 593 UCB Boulder, CO 80309-0593, USA; jennifer.kulow@colorado.edu, kevin.france@colorado.edu, robert.loyd@colorado.edu}
\altaffiltext{2}{JILA, University of Colorado and NIST, 440 UCB Boulder, CO 80309-0440, USA; jlinsky@jilau1.colorado.edu}
\altaffiltext{3}{NASA Nancy Grace Roman Fellow}

\begin{abstract}

To date, more than 750 planets have been discovered orbiting stars other than the Sun. Two sub-classes of these exoplanets, ``hot Jupiters" and their less massive counterparts ``hot Neptunes," provide a unique opportunity to study the extended atmospheres of planets outside of our solar system. We describe here the first far-ultraviolet transit study of a hot Neptune, specifically GJ436b, for which we use \emph{HST}/STIS Lyman-$\alpha$ spectra to measure stellar flux as a function of time, observing variations due to absorption from the planetary atmosphere during transit. This analysis permits us to derive information about atmospheric extent, mass-loss rate from the planet, and interactions between the star and planet. We observe an evolution of the Lyman-$\alpha$ lightcurve with a transit depth of GJ436b from $8.8\pm4.5\%$ near mid-transit, to $22.9\pm3.9\%$ $\sim2$ hours after the nominal geometric egress of the planet. Using data from the time-tag mode and considering astrophysical noise from stellar variability, we calculate a post-egress occultation of $23.7\pm4.5$\%, demonstrating that the signature is statistically significant and of greater amplitude than can be attributed to stellar fluctuations alone. The extended egress absorption indicates the probable existence of a comet-like tail trailing the exoplanet. We calculate a mass-loss rate for GJ436b in the range of $3.7\times10^6 -1.1\times10^{9}$ g s$^{-1}$, corresponding to an atmospheric lifetime of $4\times10^{11}-2\times10^{14}$ years.
\end{abstract}


\section{Introduction}\label{intro}

Since the mid-1990's, astronomers have been regularly discovering planets orbiting other stars
using the radial velocity method \citep{Mayor_1995}. Many of these planets called ``hot Jupiters" are massive ($\gtrsim50~ \mathrm{M}_\oplus \gtrsim 0.15 ~\mathrm{M}_\mathrm{Jup}$), orbit very close to their
host stars ($\le$ 0.05 AU), and have orbital periods of only a few days \citep{Seager_2010}. Less massive planets ($10-50~ \mathrm{M}_\oplus \approx 0.03-0.15~ \mathrm{M}_\mathrm{Jup} \approx 0.6-3~ \mathrm{M}_\mathrm{Nep}$), that also orbit very close to their host stars, are similarly named after their solar system
analogs, ``hot Neptunes.'' The majority of the first exoplanet detections were ``hot Jupiters'' and ``hot Neptunes'' because there is a detection bias in favor of
this type of planet. Since they are large (in mass and radius) and orbit close to their 
host stars, they are more easily detected in both radial velocity and transit searches.

In 1999, \citet{Henry_2000} discovered the first transiting exoplanet, HD209458b. Transiting exoplanets provide an opportunity 
to study the composition and structure of their atmospheres, because as 
these planets pass in front of their host star, their atmosphere blocks a portion of the 
starlight.  Spectroscopy at IR and optical wavelengths has led to the discovery of
\ion{Na}{1}, H$_2$O, CH$_4$, and CO (\citealt{Charbonneau_2002}; \citealt{Tinetti_2007}; 
\citealt{Swain_2008}; \citealt{Swain_2009}) in the atmospheres of exoplanets, shifting the 
emphasis of exoplanet studies toward detecting and characterizing their atmospheres 
\citep{Koskinen_2010}.  

While many studies utilize IR ($700\mathrm{~nm}-3~\mu$m) (e.g. \citealt{Deming_2007}) and optical ($400-700$ nm) (e.g. \citealt{Butler_2004}) wavelengths, UV ($1000-4000$ \AA) observations provide 
a unique opportunity to characterize exoplanet exospheres. Stellar FUV and EUV radiation heats, accelerates, and, when the photon energies exceed 13.6 eV, ionizes the hydrogen in the upper atmosphere \citep{Murray_2009}. When the thermal energy exceeds the gravitational potential energy ($3kT/2 > GMm/r$), the heated gas expands \citep{Lammer_2003} and likely escapes from the planet. These 
inflated atmospheres contain many atomic species, such as H, C$^+$, O, and Si$^{2+}$, that 
absorb the stellar radiation in UV resonance lines \citep{Vidal-Madjar_2004,Linsky_2010}. 
Because the envelopes of these exoplanets are inflated, the geometric area in their UV resonance lines is larger than that of molecules detected in the IR and optical, and the transit depth will be larger.
Such data permit us to understand the composition of extended exoplanet atmospheres.

\citet{Vidal-Madjar_2003} used the \ion{H}{1} Lyman-$\alpha$ 1215 \AA~emission line
of HD209458 to obtain the first evidence of an inflated exoplanetary atmosphere, 
revealing a transit depth in Lyman-$\alpha$ ($15\pm4\%$) several times that of the geometric depth 
determined from the optical transit ($1.58\pm0.18\%$). The Lyman-$\alpha$ transit depth indicates an atmosphere 
that extends beyond its Roche lobe, likely indicating atmospheric escape. Subsequent UV observations have demonstrated the existence of enlarged envelopes in the spectral lines of O, C$^+$ \citep{Vidal-Madjar_2004}, and Si$^{2+}$ \citep{Linsky_2010} around HD209458b, and
later observations identified a similar absorption spectrum from the atmosphere of HD189733b. Observations of HD189733 have been used to detect extended atmospheres 
in lines of \ion{H}{1} \citep{Lecavelier_2010}, \ion{O}{1}, and possibly \ion{C}{2} \citep{Ben-Jaffel_2013}. 
The non-detection of an extended Lyman-$\alpha$ envelope in April 2010, indicated significant 
variations in the evaporation rate of \ion{H}{1} \citep{Lecavelier_2012}.

UV transit observations are often used to infer mass-loss rates of hot Jupiters, from which 
we may learn about the evolution of exoplanets and their 
atmospheres. \citet{Vidal-Madjar_2003} used a particle simulation to limit the mass 
loss rate of HD209458b to $\ge 10^{10}$ g s$^{-1}$. \citet{Linsky_2010} performed a theoretical calculation
to find mass-loss rates in the range of $(8-40) \times 10^{10}$ g s$^{-1}$. 
 \citet{Lecavelier_2010} used their data of HD189733b and a numerical simulation to find a best fit mass-loss rate of $10^{10}$~g~s$^{-1}$. \citet{EhrenreichDesert_2011} compare the planet gravitational potential energy to the stellar X/EUV energy deposited in the atmosphere and estimate the mass-loss rate for WASP-12b to be $2.5\times10^{11}$ g s$^{-1}$.

Many theoretical models have been developed to explain the evaporation process 
and to predict the mass-loss rates for various hot Jupiters. \citet{Lammer_2003} used the heating rate from stellar X-ray and EUV radiation to estimate a mass-loss rate for HD209458b of $\sim10^{12}$ g s$^{-1}$. \citet{Yelle_2004} (later revised in \citealt{Yelle_2006}) included chemical calculations to estimate a mass-loss rate. Cooling from H$^+_3$ and ionization of H to H$^+$, reducing the amount of stellar energy available for heating, led to a lower mass-loss rate of $4.7\times10^{10}$ g s$^{-1}$. Similar calculations by \citet{Garcia_2007} determined mass-loss rates in the range $6-15\times10^{10}$ g s$^{-1}$ depending on the level of stellar activity. \citet{Holmstrom_2008} attributed a portion of the Lyman-$\alpha$ absorption to protons from the stellar wind that have been neutralized by charge exchange with hydrogen atoms in the planetary atmosphere and calculated a mass-loss rate of $7\times10^8$ g s$^{-1}$. \citet{Guo_2013} calculated a mass-loss rate of $4.3\times10^9$ g s$^{-1}$ using a two-dimensional model that assumed HD209458b is tidally locked with one side of the planet always facing the star.
\citet{Murray_2009} modeled the atmospheric escape of a theoretical hot Jupiter (similar to HD209458b) that includes realistic heating and cooling rates, ionization balance, tidal gravity, and pressure confinement by the stellar wind. They found a mass-loss rate of $2\times10^{10}$ g s$^{-1}$. In the above examples, the calculated mass-loss rates vary by several orders of magnitude depending on what physics is considered and on the values of system parameters (for example the amount of stellar EUV flux).

The manner in which the atmospheric envelope interacts with the stellar wind determines the structure of the gas around the planet. High-velocity neutral hydrogen escaping from the atmosphere trails the planet absorbing Lyman-$\alpha$ photons until the hydrogen is ionized by the stellar EUV flux or charge exchanges with stellar wind protons. The trailing material may cause the hydrogen Lyman-$\alpha$ transit to last longer than the optical occultation. The formation of a comet-like tail trailing the planet was first suggested by \citet{Schneider_1998}. Models of neutral hydrogen escaping from the hot Jupiter HD209458b and HD189733b by \citet{Bourrier_2013} support this structure. The evaporating super-Mercury exoplanet KIC 12557548b likely has a dusty comet-like tail \citep{Rappaport_2012, Budaj_2013} and the extreme hot Jupiter WASP-12b is losing sufficient mass to completely obscure its host star's emission in the cases of the \ion{Mg}{2} h and k lines \citep{Haswell_2012}.

\citet{Vidal-Madjar_2003} observed absorption out to Doppler velocities $\pm100$ km s$^{-1}$ from the Lyman-$\alpha$ line center of HD209458b, and absorption has been seen at velocities as large as -230 km s$^{-1}$ from the center of Lyman-$\alpha$ for HD189733b \citep{Lecavelier_2012}. Thermal velocities can only account for absorption out to $\sim$10 km s$^{-1}$ \citep{Murray_2009}. Models by \citet{Lecavelier_2012} suggest that stellar radiation pressure can accelerate particles up to 120 km s$^{-1}$, but an additional mechanism is necessary to explain the large observed radial velocities. Charge exchange with hot, slow ($<50-100$ km s$^{-1}$) stellar wind protons can produce the observed velocities (\citealt{Holmstrom_2008, Ekenback_2010, Tremblin_2013}). \citet{Holmstrom_2008} proposed that the absorption in neutral hydrogen at high velocities is due to charge exchange between protons from the stellar wind and planetary neutral hydrogen. The planetary hydrogen is ionized and the stellar wind proton becomes neutral hydrogen maintaining its large velocity.

\subsection{Previous Studies of GJ436b}
A promising exoplanet for UV transit observations is GJ436b, a hot Neptune, ($\mathrm{M}=0.07~\mathrm{M}_\mathrm{Jup}$), orbiting an M2 dwarf at 0.029 AU with a period of 2.6 days. GJ436b is particularly promising because it is very close to Earth, at a distance of only 10.2 pc. The properties of the system are summarized in Table~\ref{star_dat}.
GJ436b was discovered by \citet{Butler_2004} using the radial velocity method, but was later found to be transiting its host star \citep{Gillon_2007}. 

In 2010 \emph{Spitzer} observed GJ436 during several secondary transits in 6 bands from $3.6-24~\mu$m. Using a Metropolis-Hastings Markov-chain Monte Carlo (MCMC) model, \citet{Stevenson_2010} explored a wide range of parameter space to determine the best-fit compositional models. They found a high CO abundance and a deficiency of CH$_4$ relative to thermochemical equilibrium. \citet{Madhusudhan_2011} also used MCMC to confirm the overabundance of CO and CO$_2$, and a slight underabundance of H$_2$O, as compared to equilibrium chemistry with solar metallicity. They explained the observed abundances by a combination of high metallicity ($\sim10\times$ solar) and vertical mixing. 
Observations by \citet{Knutson_2014} indicate an effectively featureless
transmission spectrum, ruling out cloud-free, 
hydrogen-dominated atmosphere models. The measured 
spectrum is consistent with either a high cloud or haze layer or with a relatively hydrogen-poor atmospheric composition. \citet{Hu_2014} find that hot Neptunes, like GJ436b, are likely to have thick atmospheres that are not hydrogen dominated, but are water-rich or hydrocarbon-rich depending on their C/O ratio.
Using limited 
\emph{HST}/STIS data of the stellar Lyman-$\alpha$ flux, \citet{Ehrenreich_2011} developed numerical simulations to determine 
the transit signature of GJ436b for various assumed mass-loss rates. They predicted an 11\% transit 
depth in Lyman-$\alpha$ for a mass-loss rate of $10^{10}$ g s$^{-1}$. The analysis we have conducted is the first to place observational constraints on the mass-loss rate of GJ436b or any other Neptune-mass exoplanet.

In this paper, we present and analyze transit observations of GJ436b in Lyman-$\alpha$ observed with Space Telescope Imaging Spectrograph (STIS) on the \emph{Hubble Space Telescope} (\emph{HST}). 
In Section 2 we describe the data sets used and the reduction process, and how we created the lightcurves and velocity profiles. 
We present the lightcurve and velocity profile for Lyman-$\alpha$ in Section 3 along with the measured absorption depths. 
In Section 4 we discuss the structure of the system and calculate a range of mass-loss rates for GJ436b. 
We summarize our results in Section 5. 

\begin{deluxetable*}{lccc}
\tablecaption{Properties of GJ436 and GJ436b\label{star_dat}}
\tablewidth{0pt}
\tablehead{
\colhead{Property} & \colhead{Value} & \colhead{Reference} }
\startdata
Host Star Spectral Type 	&  M2 V								&\citet{Butler_2004}\\
Distance (pc) 			&  $10.14\substack{+0.25 \\ -0.23}$ 			&\citet{vanLeeuwen_2007}\\
$M_*/M_\odot$ 		&  $0.452\substack{+0.014\\-0.012}$		&\citet{Torres_2008}\\
$R_*/R_\odot$ 			&  $0.464\substack{+0.009\\-0.011}$		&\citet{Torres_2008}\\
$P_{orbit}$(days) 		& $2.643850\pm9\times10^{-5}$			&\citet{Pont_2009}\\
Transit Center (JD) 		& $2454279.436714\pm1.5\times10^{-5}$	&\citet{Pont_2009}\\
RA (h:m:s) 			&  +11:42:11.18 						&\citet{Zacharias_2012}\\
Dec (d:m:s) 			& +26:42:22.64  						&\citet{Zacharias_2012}\\
$R_{planet}/R_{Jup}$ 	& $0.3767\substack{+0.0082\\-0.0092}$ 		&\citet{Torres_2008}\\
$M_{planet}/M_{Jup}$ 	& $0.0727\pm0.0032$ 					&\citet{Butler_2006}\\
Semimajor axis (AU) 	& $0.02872\pm0.0048$					&\citet{Butler_2006} \\
Transit duration (hours) 	& $0.7608\pm0.012$					&\citet{Pont_2009}\\
Transit depth for $R_{planet}$ 			& $0.00696\pm0.000117$ 	&\citet{Torres_2008}\\
Escape speed from GJ436b (km s$^{-1}$) 	& 26.4 					&\\
Orbital velocity amplitude (km s$^{-1}$)	& 118 					&\\
Radial Velocity (km s$^{-1}$)			& $9.6\pm0.1$				&\citet{Nidever_2002}
\enddata
\end{deluxetable*}

\section{Observations and Data Analysis}

\begin{deluxetable}{cccrccr}
\tablecaption{Summary of Observations\label{obs}}
\tablewidth{0pt}
\tablehead{
\colhead{Data Set} & \colhead{Exp. Time (s)} & \colhead{Day} & \colhead{Start (UT)} & \colhead{Phase (hr)\tablenotemark{a,b}} 
}
\startdata
obgh0710    & 1515.145  & Dec 7, 2012     &   9:48:48 & -01:02:17.1\\
obgh0720    & 2905.121  & Dec 7, 2012     & 10:58:55 &  00:21:25.3\\
obgh0730    & 2905.193  & Dec 7, 2012     & 12:34:38 &  01:57:08.8\\
obgh0740    & 2905.172  & Dec 7, 2012     & 14:10:21 &  03:32:52.4
\enddata
\tablenotetext{a}{Phases shown are at the time of mid-exposure.}
\tablenotetext{b}{A phase of 0:00:00 corresponds to the center of the primary optical transit.}
\end{deluxetable}

\subsection{GJ436 Data\label{GJ436 Data}}
The \emph{HST}/STIS transit data of GJ436 were obtained with the G140M grating 
using a long slit with dimensions 52\arcsec$\times$0.1\arcsec. We selected a central wavelength of 1222 \AA, covering the spectral range of 1194 \AA~- 1249 \AA~with a spectral resolution of about 25 km s$^{-1}$. The data were taken with the FUV-MAMA detector using the time-tag mode.
Table~\ref{obs} summarizes all of the observations. 
Our four orbit per transit observing cadence is comparable to that used by \cite{Lecavelier_2012} for velocity resolved observations of the extended hydrogen atmosphere of the hot Jupiter HD 189733b with the same STIS grating configuration. Given an error of 2\%, the relative photometric accuracy of STIS, issues of persistence and instrumental settling are less significant when observing the 10\% UV transit signal of the exosphere compared to the $< 1\%$ near-IR molecular diagnostics of the lower atmosphere.

\begin{figure}
\epsscale{1}
\plotone{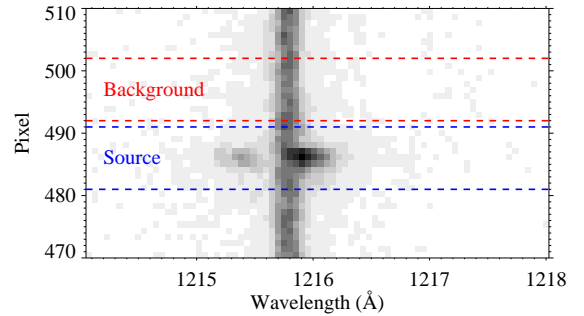}
\caption[2-D STIS data of GJ436]{
2-D STIS data of GJ436. The source ribbon used to extract the 1-D spectrum is marked in blue. The ribbon used for background subtraction is shown in red. The vertical stripe in the image is geocoronal Lyman-$\alpha$ restricted by the STIS slit.
}
\label{2d}
\end{figure}

We reduced the GJ436 
data using CALSTIS v2.40 (2012 May 23). For the final 
two exposures, the CALSTIS pipeline did not correctly 
extract the one-dimensional spectrum (the ``x1d" file) from the two-dimensional data.  
We therefore manually extracted the one-dimensional spectrum for all four exposures. 
Since in the ``x2d" file, the source was located
at pixels $481-491$, we subtracted a background ribbon at pixels $492-502$ from
the source data (see Figure~\ref{2d}). A wavelength solution was then calculated by using the reference 
wavelength, reference pixel, and dispersion from the header. Finally, we obtained fluxes 
by multiplying by the angular area covered by the pixels and an
additional scale factor (determined empirically by matching to a correctly extracted
spectrum from the ``x1d" files for the first two exposures).

\subsection{Data Analysis}
To characterize the transit, we take two complementary approaches. The first is the creation and analysis of lightcurves. This approach allows us to study the transit with higher time resolution. The lightcurve creation method is described in detail below. In the second method, we analyze the spectral difference between the pre-ingress, transit, and post-egress data. This method maintains the velocity resolution of the data. 
We also created velocity profiles for Lyman-$\alpha$. We compared the pre-ingress spectrum to the transit and post-egress spectra as a function of velocity. We also looked at the difference between the spectra (pre-ingress minus post-egress, such that absorption is a positive difference) vs. velocity. We then calculated the post-egress depth by integrating this difference spectrum.

\subsubsection{Light Curve Creation}
The data were obtained over a time period that covers portions of the exoplanet orbit before, during, and after transit. For every exposure, we calculated the orbital phase of the exoplanet (in hours) at the time of mid-exposure using the ephemeris data in Table~\ref{star_dat}, such that mid-transit occurs at phase 0 hours. 

We used the 1-D spectra extracted from the ``x2d" files, as described in Section~\ref{GJ436 Data}, and integrated the flux for different wavelength portions of the Lyman-$\alpha$ line. Avoiding the center of the Lyman-$\alpha$ line, where noise from the airglow subtraction is high, we integrated the flux from the blue and red sides of the line separately.
The wavelength range for the integration of each wing of Lyman-$\alpha$ is shown in Table~\ref{lines}. We then created lightcurves for each wing by plotting the total flux vs. orbital phase. 

We also used the ``tag" files containing the time-tag data to calculate a time series of the data in 12 minute increments. The ``tag" files contain a photon event list giving the time and detector coordinates of each recorded count. Using the known positions of the source and the background as well as the wavelength solution (from the ``x2d" files), we keep only events with detector positions that correspond to source or background counts within specified wavelength ranges. These data are then binned in time. We found that a bin size of 12 minutes is a suitable compromise between signal-to-noise and time resolution. This additional time resolution allows us to better assess the lightcurve for stellar variability \citep{Loyd_2014}. We incorporated these time-tag data in the Lyman-$\alpha$ lightcurve analysis for GJ436.

\begin{deluxetable}{lcccc}
\tablecaption{Lyman-$\alpha$ Wings\label{lines}}
\tablewidth{0pt}
\tablehead{
		& \colhead{Wavelengths}		& \colhead{Velocities} 			\\
\colhead{}& \colhead{Integrated (\AA)}	& \colhead{Integrated (km s$^{-1}$)} } 
\startdata
Blue Wing		&1214.8 -- 1215.6	&  -214.6 -- -17.3	\\
Red Wing		&1215.9 -- 1216.5	&     56.7 -- 204.7		
\enddata
\end{deluxetable}

\section{Results and Discussion}

\subsection{GJ436b Transit} 
\begin{figure}
\epsscale{1}
\plotone{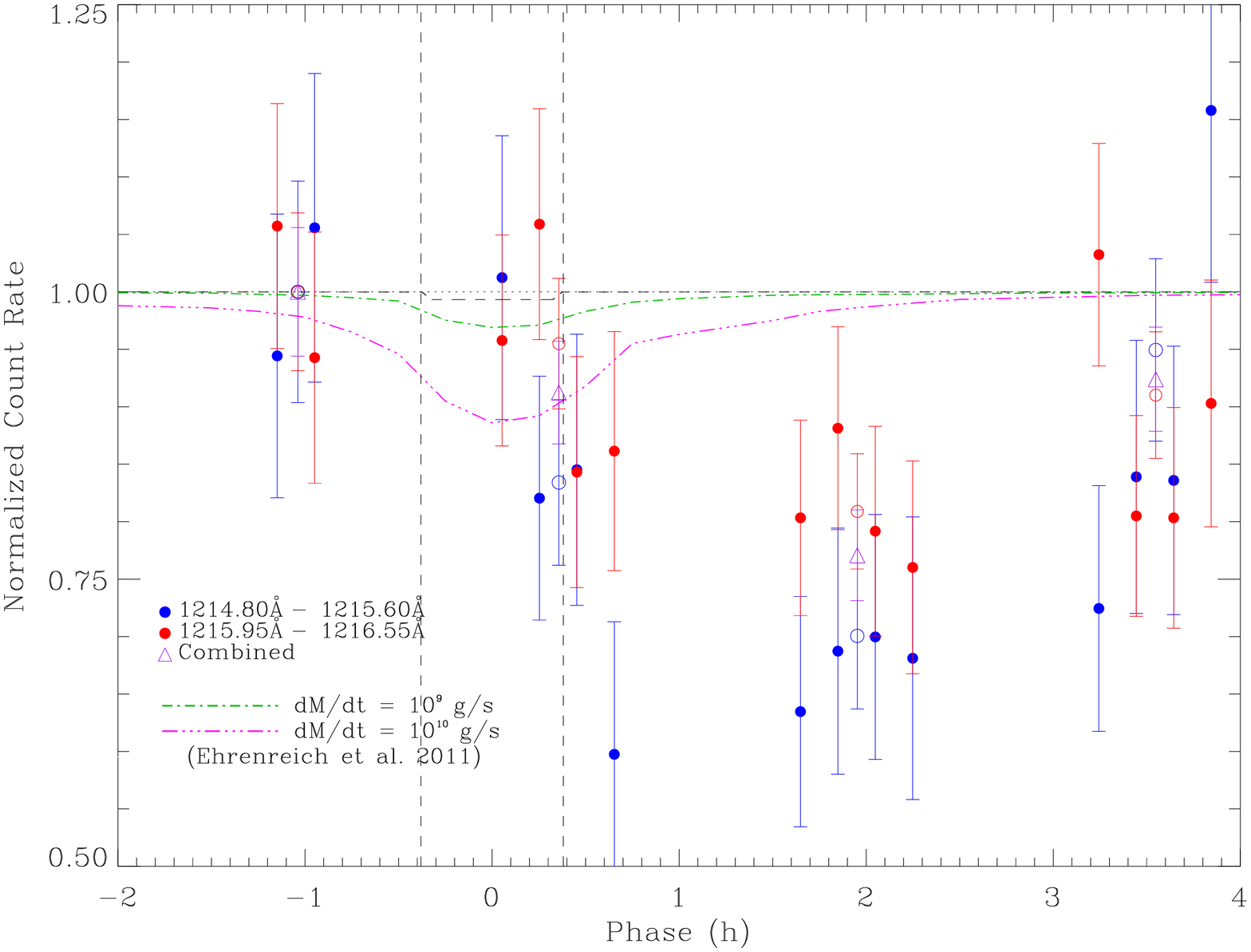}
\caption[Normalized Lyman-$\alpha$ count rates for GJ436]{
Normalized Lyman-$\alpha$ count rates for GJ436. Blue points show the flux from the 
blue wing of Lyman-$\alpha$ and red points show the flux from the red wing. 
Filled points are calculated from the time-tag data, while the open circles are 
from the entire exposure. Error bars indicate $\pm1\sigma$. The black dotted 
line indicates the normalized flux level, 
while the black dashed line shows the transit curve as calculated from the optical 
transit parameters. The green and pink lines shows the predicted transit signature 
of GJ436b as calculated by \citet{Ehrenreich_2011} for mass-loss rates of $10^9$ 
and $10^{10}$ g s$^{-1}$ respectively. Vertical dashed lines indicate first and fourth contacts.
}
\label{LyA_int}
\end{figure}

\begin{figure}
\plotone{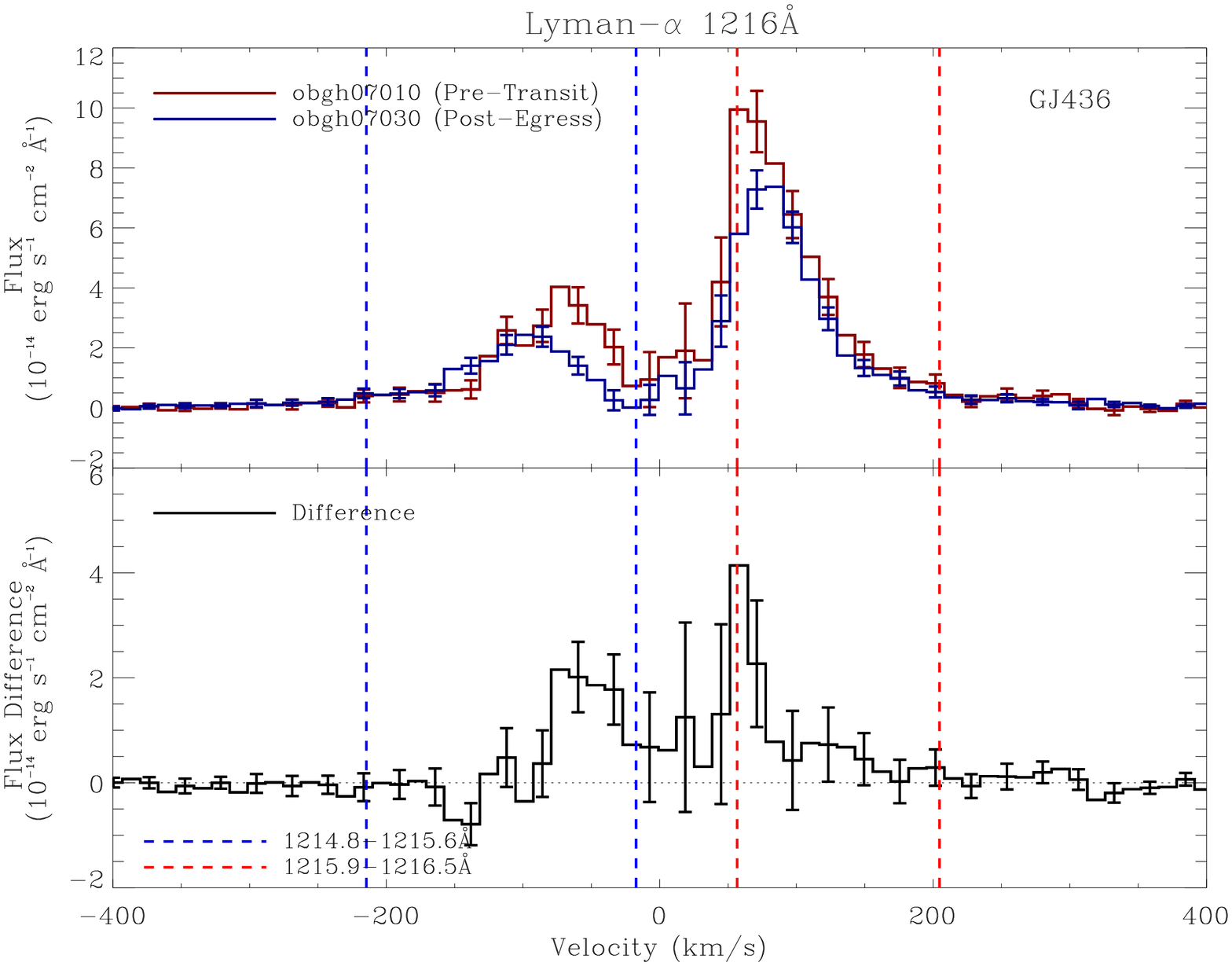}
\caption[Lyman-$\alpha$ velocity profile for GJ436]{
Lyman-$\alpha$ velocity profile for GJ436. The top panel compares the
pre-ingress spectrum, in dark red, to the post-egress spectrum, in dark blue, and the bottom panel 
shows the difference between these spectra, pre-ingress minus post-egress. Error bars indicate 
$\pm1\sigma$. The regions of integrated flux are 
also shown; the blue wing is between the blue dashed lines and the red wing is between 
the red dashed lines. During the post-egress time interval, we find an occultation depth of $29.9\pm8.3\%$ in the Lyman-$\alpha$ blue wing, $19.1\pm5.8\%$ in the Lyman-$\alpha$ red wing, and $22.9\pm4.8\%$ in the combined wings of the Lyman-$\alpha$ line.
}
\label{diff}
\end{figure}

We show the Lyman-$\alpha$ lightcurve for GJ436 in Figure \ref{LyA_int} and the Lyman-$\alpha$ velocity profile in Figure \ref{diff}.
Table~\ref{GJ436_tr_dep} shows the transit depths extracted from these figures. The transit depths are identical for the two procedures (because the reference spectrum and the integration method are the same), however, the spectral analysis results in somewhat larger errors, due to the extra step of subtracting prior to normalizing and integrating. We will use the higher time resolution lightcurve data to analyze the transit depth. For GJ436b we see a mid-transit depth of $16.6\pm7.2\%$ in the blue wing and $4.5\pm5.7\%$ in the red wing. This corresponds to an occulting disk of 5.0 R$_\mathrm{p}$ and 2.6 R$_\mathrm{p}$ respectively, both smaller than the Roche lobe radius of 6.1 R$_\mathrm{p}$. When both wings are combined, the mid-transit depth is $8.8\pm4.5\%$, corresponding to an equivalent opaque occulting disk of 3.6 R$_\mathrm{p}$. The asymmetry in the absorption can be explained by charge exchange of the stellar wind with the atmosphere. As viewed from Earth, only the stellar wind traveling towards us can be observed, causing excess absorption blue-ward of line center. This type of asymmetry, with more absorption in the blue wing, is also seen in the Lyman-$\alpha$ absorption during transit from HD209458b \citep{Vidal-Madjar_2003} and HD189733b \citep{Lecavelier_2012}.

Interestingly, the Lyman-$\alpha$ transit extends much later in phase than the optical transit. We examine these data for the possibility of extended egress, finding post-egress depths of $29.9\pm6.4\%$ in the blue wing, $19.1\pm5.0\%$ in the red wing, and $22.9\pm3.9\%$ for both wings combined $\sim2$ hours after mid-transit and about 1.5 hours after fourth contact. The time-tag data points are presented in Table~\ref{GJ436_tr_dep_tt}. These data corroborate the detection of both the transit and 
extended egress, as three of the four transit data points and all of the post-egress data points show a transit detection, although large variations in the blue-wing time-tag data are observed near mid-transit. 

\begin{deluxetable}{lcc}
\tablecaption{Transit Depths for GJ436 Full Exposures\label{GJ436_tr_dep}}
\tablewidth{0pt}
\tablehead{
& \colhead{Transit Depth from}    & \colhead{Transit Depth}   \\
\colhead{Species}& \colhead{Difference Spectrum}& \colhead{from Lightcurve} }
\startdata
Lyman-$\alpha$ Blue Wing mid-transit    & $16.6\pm8.2$\%  &  $16.6\pm7.2$\%  \\
Lyman-$\alpha$ Red Wing  mid-transit    & $4.5\pm5.9$\%  &  $4.5\pm5.7$\%  \\
Lyman-$\alpha$ coadded    mid-transit    & $8.8\pm4.8$\%  &  $8.8\pm4.5$\%  \\ \\
Lyman-$\alpha$ Blue Wing post-egress & $29.9\pm8.3$\%  &  $29.9\pm6.4$\%  \\
Lyman-$\alpha$ Red Wing  post-egress & $19.1\pm5.8$\%  &  $19.1\pm5.0$\%  \\
Lyman-$\alpha$ coadded    post-egress & $22.9\pm4.8$\%  &  $22.9\pm3.9$\%  
\enddata
\end{deluxetable}

\begin{deluxetable}{lcc}
\tablecaption{Transit Depths for GJ436 Time-Tag Points\label{GJ436_tr_dep_tt}}
\tablewidth{0pt}
\tablehead{
& \colhead{Transit Depth in}    & \colhead{Transit Depth in}   \\
\colhead{Phase (hr)}&\colhead{Blue Wing}&  \colhead{Red Wing} }
\startdata
-1:08:54.7	& $ 5.6\pm12.3$\%	& $-5.7\pm10.7$\% \\
-0:56:54.7	& $-5.6\pm13.4$\%	& $ 5.7\pm10.9$\% \\
 0:03:12.3	& $-1.2\pm12.4$\%	& $ 4.2\pm9.2$\% \\
 0:15:12.3	& $ 18.0\pm10.6$\%	& $-5.9\pm10.0$\% \\
 0:27:12.3	& $ 15.5\pm11.8$\%	& $15.7\pm10.1$\% \\
 0:39:12.3	& $ 40.2\pm11.5$\%	& $13.9\pm10.4$\% \\
 1:38:55.3	& $ 36.5\pm10.0$\%	& $19.7\pm8.5$\% \\
 1:50:55.3	& $ 31.3\pm10.7$\%	& $11.9\pm8.8$\% \\
 2:02:55.3	& $ 30.0\pm10.7$\%	& $20.8\pm9.1$\% \\
 2:14:38.5	& $ 31.9\pm12.3$\%	& $24.0\pm9.3$\% \\
 3:14:38.4	& $ 27.6\pm10.7$\%	& $-3.2\pm9.7$\% \\
 3:26:38.4	& $ 16.1\pm11.9$\%	& $19.5\pm8.7$\% \\
 3:38:38.4	& $ 16.4\pm11.7$\%	& $19.7\pm9.6$\% \\
 3:50:38.4	& $-15.8\pm15.0$\%	& $ 9.7\pm10.7$\% 
\enddata
\end{deluxetable}

\subsection{Extended Egress in GJ436\label{egress}}
To determine whether or not the extended egress is real, we 
consider the time-tag data. We form the null hypothesis that the data are Gaussian distributed 
and the apparent occultation is random noise. We use the average of the error bars on the time-tag 
data as the standard deviation for the Gaussian distributions ($\sigma_{blue}=0.1092, 
\sigma_{red}=0.08930$), both with a mean of unity. Assuming these distributions, we determined the 
probability of finding four consecutive points lower than the highest point in the set of four at the deepest 
transit depth, which is located at phase $\sim$2 hours ($x_{blue}=0.3003, x_{red}=0.1187$ below the 
mean). 
We did this by randomly picking 14 points from the specified Gaussian distribution and counting
how many trials out of $10^7$ had four consecutive points outside the requisite range. For the 
parameters determined for the blue wing, none of the $10^7$ trials had four consecutive points. 
For the red wing distribution, we found a probability of 9.31$\pm0.76\times10^{-5}$ to randomly produce the result. We therefore conclude that the deep, extended egress signal seen in Figure \ref{LyA_int} is real and not due to random statistical variations.

\subsection{Stellar Variability of GJ436\label{variability}}
While we have determined that the post-egress detection is not due to statistical noise, we have not yet 
considered whether the drop in flux could be due to stellar variability, as opposed to atmospheric absorption from GJ436b. To address this issue, we look at a resonance line from \ion{N}{5}, an ion that we do not expect to find in the atmosphere of the exoplanet. For the \ion{N}{5} time-tag data points we find $\mathrm{RMS}=0.2765$, while the average value of the $1\sigma$ error bar for those points is 
0.4010. We conclude that the signal to noise is too low for the \ion{N}{5} doublet to be 
a suitable tracer of stellar variability. Similarly, there was not enough flux in the \ion{Si}{3} line to be measurable above the noise.

Instead we look to the literature to assess the potential 
magnitude of stellar variability. \citet{Loyd_2014} studied time variability in the \ion{C}{2}, \ion{Si}{3}, and \ion{Si}{4} resonance lines of 38 cool stars, including GJ436. They did not attempt to characterize Lyman-$\alpha$ line variability because geocoronal airglow cannot be removed reliably from their COS data. Instead we use their \ion{C}{2} emission line variability, because \ion{C}{2} has a similar formation temperature to Lyman-$\alpha$ (T$_{form}\approx(1-3)\times10^4$ K, \citealt{Dere_2009}). \citet{Loyd_2014} found that the mean-normalized chromospheric \ion{C}{2} line variability, excluding flare periods, in GJ436 is  $0.20^{0.09}_{0.12}$ on 60 second timescales. Our post-egress data cover a 48 minute time span. Assuming that the stellar variability is uncorrelated over time, the noise associated with stellar fluctuations is estimated to be $0.20/\sqrt{48}=0.0289$. Combining the 8 post-egress time-tag data points and the photon noise with the upper limit on the noise expected from chromospheric variability, we find a 23.7\% occultation with an uncertainty of 4.5\%. From this we conclude that the post-egress detection is very likely real. Future observations over several transit cycles would be very valuable.

\section{GJ436 System}

\subsection{Structure\label{struct}}
Some models of the interaction between escaping gas from the exoplanet's atmosphere and the stellar wind predict a comet-like tail extending behind the planet (\citealt{Schneider_1998, Bourrier_2013}). 
The Lyman-$\alpha$ data support this 
type of structure in the GJ436 system. A large, inflated cloud of gas trailing the planet could explain  
an occultation deeper than that observed in the optical and post-egress 
absorption long after optical transit. 

In order to simulate the structure of the exoplanetary gas cloud at the orbit of GJ436b, we attenuate the
pre-ingress profile by a column of hydrogen and deuterium to match the post-egress spectrum. We search for a $\chi^2$
best fit neutral hydrogen column ([log(N$_{HI}$)] ranging from 14.0 to 19.0 in step sizes
of 0.2 dex) for a grid of stellar covering fraction (assuming uniform optical thickness of the gas over the area covered and uniform emission of Lyman-$\alpha$ from the stellar disk) vs. Doppler $b$-value, $b=(\frac{2kT}{m_H}+v_{turb}^2)^{1/2}$. We assume a fixed D/H ratio 
of $1.5\times10^{-5}$ \citep{Linsky_2006}. For the range of tested covering fractions and $b$-values, we find two parameter regimes that provide a reasonable fit to the data. The first is a low covering fraction,
high N$_{HI}$ model. The second is a high covering fraction low N$_{HI}$ model. We expect 
the second regime to be more physically plausible. The low covering fraction models require N$_{HI}\sim10^{19}$ cm$^{-2}$. Assuming that the comet-like tail extends to a few times the orbital radius, $\sim0.1$ AU in the line of sight to the star, we find a number density $\mathrm{n}_H =\sim\mathrm{few} \times10^{7}\mathrm{cm}^{-3}$. This density is far above the estimates ($\sim3\times10^5$ cm$^{-3}$) from comet-like tail models \citep{Bourrier_2013}. A column density of N$_{HI}=10^{15}$ cm$^{-2}$, typical of the high covering fraction regime, converts to a number density $\mathrm{n}_{HI} =\sim\mathrm{few} \times10^{4}\mathrm{cm}^{-3}$.

\begin{figure}
\plotone{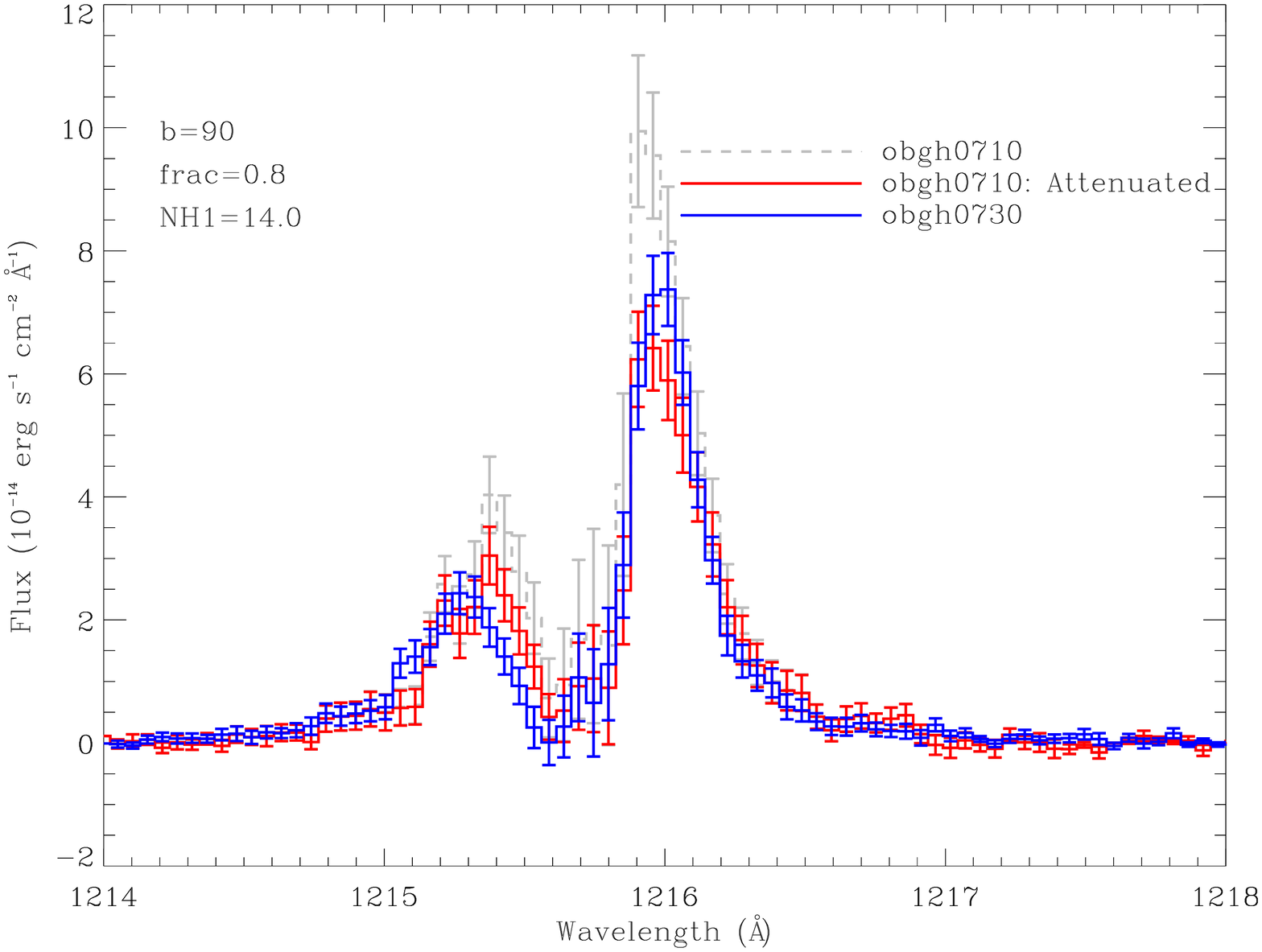}
\caption[Lyman-$\alpha$ Profile Attenuation]{
Lyman-$\alpha$ profile attenuation. The gray curve shows the Lyman-$\alpha$ 
profile pre-ingress, and the blue curve shows the profile observed post-egress. The red curve 
shows the pre-ingress profile attenuated by a cloud of hydrogen and deuterium gas 
with a Doppler $b$-value of 90 km s$^{-1}$, a covering fraction 0.8, and a $\chi^2$ best fit 
column density of N$_\mathrm{H}=10^{14.0}$ cm$^{-2}$. Error bars indicate $\pm1\sigma$.
}
\label{LyA_att}
\end{figure}

Taking the high covering fraction and low N$_{HI}$ case as more plausible, acceptable fits to the post-egress Lyman-$\alpha$ spectrum require the Doppler $b$-parameter
to be between 60 and 120 km s$^{-1}$. We require $b$-values in this range to fit the shape of the post-egress Lyman-$\alpha$ profile. For a typical atmospheric temperature of 10$^4$ K, the thermal width is 13 km s$^{-1}$, requiring a superthermal velocity component of $\sim50$ to 110 km s$^{-1}$ to explain such high $b$-values. Charge exchange between stellar wind protons and planetary wind neutral H atoms will lead to neutral H atoms with high velocities. For a representative covering fraction of 0.8, we can then limit log(N$_{HI}$) to $14-16$. Figure~\ref{LyA_att} shows an example best fit attenuation profile with parameters within this range. These ranges are shown in Table~\ref{limits}.

To achieve a covering fraction of 0.8 requires a cloud of $R\approx10~R_{planet}$. Using the column 
density and $b$ (to calculate the cross-section at the line core, $\sigma=\frac{\sqrt{\pi}e^2}{m_ec}\frac{f\lambda_0}{b}$), we can estimate the line center optical depth of the extended neutral hydrogen atmosphere ($\tau_0=N_H\sigma$). For the ranges of N$_H$ and $b$, we find $\tau_0=0.63-130$. \citet{Koskinen_2010} predict the hydrogen cloud to be optically thick in the line wings, indicating that our models with the lowest columns and higher $b$s (which give the higher $\tau_\nu$s) are more physically representative. 

\begin{deluxetable}{lc}
\tablecaption{Cloud Properties\label{limits}}
\tablewidth{0pt}
\tablehead{
& \colhead{Range}	} 
\startdata
Covering fraction		& $0.7-0.9$	\\
$b$ (km s$^{-1}$)		& $60-120$	\\
log(N$_{HI}$[cm$^{-2}$])	& $14-16$	\\
$\tau_0$				& $0.63-130$	
\enddata
\end{deluxetable}

\subsection{GJ436b Mass-Loss Rate}
\subsubsection{Spherical Mass-Loss}
\begin{figure}
\plotone{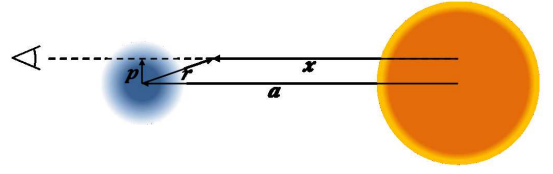}
\caption[Geometry for mass-loss calculation]{
Geometry for mass-loss calculation. We consider a line of sight (LOS) toward the center of the star that passes through the planet's atmosphere at a distance $p$ from the center of the planet, where $x$ is measured along the LOS, $r$ is measured from the center of the planet, and $a$ is the distance between the star and planet.}
\label{ml_fig}
\end{figure}

We first calculate the mass-loss rate for 
GJ436b by assuming a spherically symmetric envelope around the planet that blocks a portion of the stellar surface. We consider a line of sight (LOS) toward the center of the star that passes through the exoplanet's atmosphere a distance $p$ from the center of the planet. We measure $x$ along the LOS and $r$ along from the center of the planet (see Figure~\ref{ml_fig}). 
For a spherical outflow with a constant mass-loss rate, the mass flux in neutral hydrogen from the planet is 
\begin{equation}\label{mlr_n}
\dot{M}_{HI} =4\pi r^2 m_H v n_{HI}(r),
\end{equation}
where $v$ is the outflow velocity at $r$.
The optical depth at the line center along this LOS is
\begin{equation}\label{tau_n}
\tau_0 =\sigma \int_0^{\infty} n_{HI}(x)dx,
\end{equation}
where $n_{HI}$ is the number density of neutral hydrogen, $\sigma$ is the absorption cross section.
Combining Equations \ref{mlr_n} and \ref{tau_n} yields
\begin{equation}\label{ml_int}
\tau_0=\frac{\sigma\dot{M}_{HI}}{4\pi m_{H}}\int_0^\infty \frac{dx}{vr^2} \Rightarrow \dot{M}_{HI}=\frac{4\pi m_H \tau_0}{\sigma} \left[ \int_0^\infty \frac{dx}{vr^2} \right]^{-1}
\end{equation}
 In the integrand, the $r^{-2}$ factor gives the highest weight to $v$ values where the LOS passes closest to the planet (smallest $r$). Thus, for an order-of-magnitude calculation $v$ may be taken out of the integral and replaced with a value representative of that expected near the radius at closest approach, $r=p$, between the LOS and planet. (This assumes $v$ does not drop precipitously at large $r$, consistent with the models of \citet{Murray_2009}.) Bringing $v$ out of the integral, we are able to find an analytical solution. From Figure~\ref{ml_fig} we can use $r^2=(x-a)^2+p^2$, where $a$ is the star-planet distance, to evaluate the integral in Equation~\ref{ml_int}.
\begin{equation}\label{integ}
\int_0^\infty \frac{dx}{r^2} = \int_0^\infty \frac{dx}{(x-a)^2+p^2} = \frac{\pi}{2p}+\frac{1}{p}\arctan\left(\frac{a}{p}\right).
\end{equation}
Since only LOSs that intersect the stellar disk, $p<R_{\star}$, are of interest, and for GJ436 $a/R_\star=13.34$, $\arctan(a/p)$ may be taken to be $\pi/2$ to good accuracy, so that the integral in Equation~\ref{integ} simplifies to $\pi/p$. Thus 
\begin{equation}\label{M_dot}
\dot{M}_{HI}\approx\frac{4pvm_H\tau_0}{\sigma_0}
\end{equation}
The cross section at line center is given by
\begin{equation}
\sigma_0=\frac{\sqrt{\pi}e^2}{m_ec\Delta\nu_D}f
\end{equation}
 $f = 0.4161$ is the oscillator strength, $m_e$ is the electron mass, and $\Delta\nu_D$ is the Doppler width given by
\begin{equation}
\nonumber
\Delta \nu_D=\frac{\nu_0}{c} \left( \frac{2kT}{m_H}+v_{turb}^2 \right)^{1/2}=\frac{b}{\lambda_0} . 
\end{equation}
Given the range of $b$-values determined in Section~\ref{struct} ($b=60-120$ km s$^{-1}$), we limit the cross section at Lyman-$\alpha$ to $\sigma_0(\mathrm{HI})=6.3\times10^{-15}-1.3\times10^{-14}$.

Assuming the absorption is due to an optically thin cloud covering the star, we consider annuli of area $2\pi p dp$ centered on the planet. Each annulus absorbs a fraction of the light from the stellar disk equal to
\begin{equation}
\frac{dF_\star}{F_\star}=\frac{2\pi p dp (1-e^{-\tau_0})}{\pi R_\star^2}=\frac{2}{R_\star^2}p(1-e^{-\tau_0})dp,
\end{equation}
where $\tau_0=\tau_0(p)$ is the optical depth at line center along the LOS passing through the annulus of radius $p$. The observed transit depth ($\delta$) is then
\begin{equation}\label{tr_dep}
\delta=\int_{R_p}^{R_\star}\frac{dF}{F}=\frac{2}{R_\star^2}\int_{R_p}^{R_\star}p(1-e^{-\tau_0})dp.
\end{equation}
According to \citet{Murray_2009}, $\tau_0<0.1$ for $p\gtrsim1.5R_p$, so using the small $\tau$ approximation, Equation~\ref{tr_dep} becomes
\begin{equation}
\delta=\frac{2}{R_\star^2}\int_{R_p}^{R_\star}p\tau_0dp.
\end{equation}
Using Equation~\ref{M_dot} this becomes
\begin{equation}
\delta\approx\frac{2}{R_\star^2}\frac{\dot{M}\sigma_0}{4m}\int_{R_p}^{R_\star}\frac{dp}{v}
\approx\frac{2}{R_\star^2}\frac{\dot{M}\sigma_0}{4mv}(R_\star-R_p)
\approx\frac{2}{R_\star}\frac{\dot{M}\sigma_0}{4mv}.
\end{equation}
While \cite{Murray_2009} predict $v$ to vary by a factor of a few over the range of $p$ considered here, we have approximated $v$ to be constant with $p$. We have also neglected the $R_p \ll R_\star$ term.
Solving for the mass loss rate, we get
\begin{equation}
\dot{M}\approx\frac{2\delta R_\star m v}{\sigma_0}
\end{equation}
Following \citet{Murray_2009}, we adopt $v=10$ km s$^{-1}$, approximately equal to the thermal velocity of hydrogen at 10$^4$ K. We use the observed mid-transit depth $\delta=0.088\pm0.045$. Thus, these observations, 
for our range of $b$-values, bound the mass-loss rate in $neutral$ hydrogen to $
\dot{M}_{HI}=3.7 \times 10^5 - 2.3 \times 10^6 \mathrm{~g~s}^{-1}$.

The mass-loss rate depends linearly on the $b$-value. So, if the $b$-value at the exobase, where the wind is launched, differs from $b$-values we have calculated, the mass-loss rate would differ correspondingly.
Because the planet is so close to its host star, the EUV flux and charge exchange with the stellar wind will ionize most of the escaping hydrogen and only a small fraction will be neutral in the outer atmosphere. \citet{Koskinen_2013} model the H and H$^+$ density in the atmosphere of HD209458b as a function of planetary radius. In the upper atmosphere ($R \sim 5R_p$) the neutral fraction is $\sim0.1$. Assuming similar ionization conditions in the extended atmosphere of GJ436b, we correct for this neutral fraction, and calculate a total mass-loss rate 
\begin{equation}
\dot{M}=3.7 \times 10^6 - 2.3 \times 10^{7} \mathrm{~g~s}^{-1}.
\end{equation}

Assuming the structure of GJ436b is similar to that of Neptune, and the atmosphere comprises 5-15\% of the mass of the planet \citep{Guillot_1999}, and that the mass-loss rate is constant in time, this range of mass-loss rates give a range of atmospheric lifetimes of $9.5\times10^{12}-1.8\times10^{14}$ years, indicating that the atmosphere is stable over the lifetime of the star.

\subsubsection{EUV Heating to PdV Work}
We also calculate the mass-loss rate following the analytical argument of \citet{Murray_2009}. We first calculate the amount of stellar EUV flux available to heat the atmosphere,
\begin{equation}
E_{heat}=\epsilon \pi F_{EUV} R_p^2~[\mathrm{erg~s}^{-1}].
\end{equation}
Here $\epsilon$ is an efficiency factor, $F_{EUV}$ (in erg cm$^{-2}$ s$^{-1}$) is the stellar flux at the orbit of the planet from $300-912$ \AA~(where the photoionization cross-section for hydrogen is largest; \citealt{Murray_2009}), and  $R_p$ is the planetary radius ($R_p=0.38~R_{Jup}$). Using the model scaling relations of \citet{Linsky_2014}, we estimated the EUV flux based on the reconstructed Lyman-$\alpha$ luminosity \citep{France_2013}. These calculated EUV luminosities are shown in Table~\ref{EUV_lum}.
We then consider the PdV work required to liberate a unit mass from the gravitational well of the planet:
\begin{equation}
\frac{P\Delta V}{\rho R_p^2 H}\sim\frac{P R_p^3}{\rho R_p^2 H}\sim\frac{\rho g H R_p^3}{\rho R_p^2 H}\sim\frac{G M_p}{R_p}.
\end{equation}
The mass-loss rate is then the ratio of the heat available to the work required to lift out mass:
\begin{equation}
\dot{M}=\frac{\epsilon \pi F_{EUV} R_p^2}{G M_p/R_p} = \frac{\epsilon \pi F_{EUV}R_p^3}{G M_p} =1.1\times10^9~\mathrm{g~s^{-1},}
\end{equation}
corresponding to a lifetime of $4\times10^{11}$ years. Here we have assumed an efficiency $\epsilon=0.3$ and extrapolated an EUV flux $F_{EUV}=607$ erg cm$^{-2}$ s$^{-1}$ from the reconstructed intrinsic Lyman-$\alpha$ flux. This argument is valid only if we are in the EUV driven, as opposed to X-ray driven, evaporation regime. According to \citet{Owen_2012} for a Neptune mass planet with a density of 1 g cm$^{-3}$ at a separation of 0.025 AU (for GJ436b: mass = 1.35 M$_{\mathrm{Nep}}$, density = 1.69 g cm$^{-3}$, separation = 0.029 AU) the critical X-ray luminosity at which the planetary wind will transition from X-ray driven to EUV driven is $\sim8\times10^{28}$ erg s$^{-1}$. \citet{Kashyap_2008} measure the the X-ray luminosity of GJ436 to be $1.4\times10^{27}$ erg s$^{-1}$, two orders of magnitude lower than the transition value, placing GJ436b well into the EUV driven evaporation regime.

\begin{deluxetable}{rc}
\tablecaption{EUV Luminosity of GJ436\label{EUV_lum}}
\tablewidth{0pt}
\tablehead{
\colhead{Band} & \colhead{Luminosity [erg s$^{-1}$]}	} 
\startdata
\textless100\AA\tablenotemark{a}	&  $1.45\times10^{27}$ \\
100-200\AA	&  $1.44\times10^{27}$ \\
200-300\AA	&  $1.26\times10^{27}$ \\
300-400\AA	&  $1.12\times10^{27}$ \\
400-500\AA	&  $2.48\times10^{25}$ \\
500-600\AA	&  $4.47\times10^{25}$ \\
600-700\AA	&  $5.82\times10^{25}$ \\
700-800\AA	&  $6.98\times10^{25}$ \\
800-912\AA	&  $9.41\times10^{25}$ 
\enddata
\tablenotetext{a}{ 5$-$124 \AA~luminosity from \citet{Kashyap_2008}.}
\end{deluxetable}

These two methods give results that differ by 2 orders of magnitude. This is not surprising given that mass-loss calculations for HD209458 can vary by 4 orders of magnitude depending on method. Our range of mass-loss values are roughly consistent with the models of \citet{Ehrenreich_2011}. Their model transit curves can be seen in Figure~\ref{LyA_int}. They predict a transit depth of 11\% for a mass-loss rate of $10^{10}$ g s$^{-1}$. Their calculation of the mass-loss rate was based on the measured stellar X-ray ($5-124$ \AA) luminosity, log($L_{X}$ [erg s$^{-1}$])$=26.85\substack{+0.65\\-0.89}$ \citep{Hunsch_1999}. However, this X-ray luminosity arises in the stellar corona, and may not be representative of the majority of the longer wavelength EUV flux from GJ436, which is likely dominated by Lyman continuum emission from the transition region and upper chromosphere \citep{Linsky_2014}. We find EUV luminosities lower than \citet{Ehrenreich_2011}, especially at wavelengths where the photoionization cross-section for hydrogen is largest ($300-900$ \AA; \citealt{Murray_2009}). Their calculation also depends linearly on the heating efficiency ($\eta$, similar to $\epsilon$ above) in the atmosphere, a parameter that is not well constrained. For $\eta=0.15$ they calculate $\dot{M}=1.60\times10^9$ g s$^{-1}$, but this mass-loss rate can range between $1.07\times10^8$ g s$^{-1}$ and $1.07\times10^{10}$ g s$^{-1}$ for $\eta$ between 0.01 and 1. Uncertainties in the neutral fraction that we use in our first calculation, $\epsilon$ in our second calculation, and the parameters of the \citet{Ehrenreich_2011} calculation leave a wide range for the possible mass-loss rate of GJ436b.

\section{Summary}
We have analyzed new observations of GJ436. We used \emph{HST}/STIS data to detect and characterize 
the extended atmosphere of GJ436b for the first time. We detected $8.8\pm4.5\%$ absorption in the Lyman-$\alpha$ line at mid-transit, and used this transit depth to calculate a mass-loss rate in the range $3.7\times10^6-1.1\times10^{9}$ g s$^{-1}$, corresponding to an atmospheric lifetime of $4\times10^{11}-2\times10^{14}$ years. We also detected strong absorption after the optical transit with a depth of $23.7\pm4.5\%$. We confirmed that this extended egress is not a statistical fluctuation, and showed that it is unlikely to be due to stellar variability; the most likely explanation is that GJ436b is trailed by a comet-like tail of neutral hydrogen.

This material is based upon work supported by the National Science Foundation 
Graduate Research Fellowship Program under Grant No. DGE 1144083. The data presented here were
obtained as part of \emph{HST} Observing program \#12034. KF acknowledges support through
a NASA Nancy Grace Roman Fellowship during this work.

\bibliography{ms}{}

\end{document}